\documentclass[aps,prb,twocolumn,superscriptaddress,showpacs]{revtex4-1}

\usepackage{amsmath,amssymb,amsfonts}
\usepackage{graphicx}
\usepackage{hyperref}

\begin{document}
\title{Theory of a quantum critical phenomenon in a topological insulator: 
(3+1) dimensional quantum electrodynamics in solids}
\author{Hiroki Isobe}
\affiliation{Department of Applied Physics, University of Tokyo, Tokyo 113-8656, Japan}
\author{Naoto Nagaosa}
\affiliation{Department of Applied Physics, University of Tokyo, Tokyo 113-8656, Japan}
\affiliation{Cross Correlated Materials Research Group (CMRG) and Correlated
Electron Research Group (CERG), ASI, RIKEN, Wako, Saitama 351-0198, Japan}
\date{\today}

\begin{abstract}
We study theoretically the quantum critical phenomenon of the
phase transition between the trivial insulator and the topological 
insulator in (3+1)-dimensions, which is described by a Dirac 
fermion coupled to the electromagnetic field.
The renormalization group (RG) equations for the running 
coupling constant $\alpha$, the speed of light $c$, and electron
$v$ are derived. The almost exact analytic solutions to 
these RG equations are obtained to reveal that (i) $c$ 
and $v$ approach to the common value 
with combination $c^2 v$ being almost unrenormalized, (ii) 
the RG flow of $\alpha$ is the same as 
that of usual QED with $c^3$ being replaced by $c^2 v$, and 
(iii) there are two crossover momentum/energy scales separating
three regions of different scaling behaviors. 
The dielectric and magnetic susceptibilities,
 angle-resolved photoemission
spectroscopy (ARPES), and the behavior of the gap
are discussed from this viewpoint.
\end{abstract}
\maketitle

\section{Introduction}
In solids, the electronic states are 
described by the Bloch wavefunctions with the 
energy dispersion $\varepsilon_n(\vec{k})$ where 
$n$ being the band index and $\vec{k}$ the 
crystal momentum. The velocity of electrons given by 
$\vec{v}_n (\vec{k}) =
\partial \varepsilon_n(\vec{k})/\partial \vec{k}$ 
is usually much smaller than that of light $c$. 
Therefore, the Lorentz invariance is terribly broken and hence
many of the beautiful results in quantum electrodynamics (QED)~\cite{Peskin}  
are not applicable to the
Bloch electrons in solids. 
The smallness of the factor $v_n(\vec{k})/c \ll 1$ 
naturally leads to the gauge choice (i.e., Coulomb gauge) 
where the scalar potential gives the Coulomb interaction 
without retardation while the electron-electron 
interaction through the transverse part of the
vector potential $\vec{A}$ is often neglected.
The latter is often treated as the external electromagnetic
field for the probe of the electromagnetic
response of the system. This gauge choice is regarded as
the ``physical gauge.'' For example,
one can discuss the physical meaning
of the Green's function $G(\vec{k}, \omega)$
in this gauge where the quasi-particle 
corresponds its pole structure. 
Angle-resolved photoemission spectroscopy 
(ARPES) is also formulated in this gauge [i.e.,
ARPES intensity is proportional to
the electron spectrum function  $-\mathrm{Im} G(\vec{k}, \omega)$]~\cite{Shen}. 
 
While the nonrelativistic quantum mechanics is basically
justified for the electrons in solids, there are some cases
where the Dirac fermions appear in the electronic band structure.
A representative case is graphene, a two-dimensional 
sheet of carbon network with hexagonal lattice, where the
$2\times 2$ Dirac spectrum near $K$ and $K'$ points describes the
low-energy physics~\cite{neto1}. Another example is Bi, which is described by 
$4\times 4$ Dirac fermions and shows the enhanced orbital diamagnetism~\cite{Fuseya}.
Recent advances in this field are the 
discovery of the topological insulator and 
its associated quantum phase transition~\cite{TI1,TI2}. The relativistic spin-orbit
interaction (SOI) rearrange the spin states to yield the 
``twist'' of the Bloch wavefunctions in the first Brillouin zone.
This twist is characterized by the $Z_2$ topological integer.
In general, topological integers can change only discontinuously
when the gap closes, which can be described by the local
Hamiltonian in $k$ space. When the inversion symmetry exists,
the effective Hamiltonian near this quantum phase transition is
the Dirac Hamiltonian expanded around the time-reversal 
invariant momentum (TRIM) $\vec{k}_0$ ($\vec{k}_0$
is equivalent to $-\vec{k}_0$). In this case, the
orbitals and spins are coupled to form the $4\times 4$
Dirac Hamiltonian and the sign change 
of the mass $m$ corresponds to the 
quantum phase transition between 
trivial insulator and topological insulator.
This story is actually realized in the materials such 
as BiTl(S$_{1-x}$Se$_x$)$_2$ by changing the concentration $x$~\cite{Hasan,Sato}.  

The effects of the electron-electron interaction on the Dirac
electrons are also extensively studied~\cite{neto2,Gonzalez,Chakravarty,Vishwanath}. 
For graphene, it has been revealed that the electron speed $v$ is renormalized 
to increase
logarithmically by the long-range Coulomb interaction, while the coupling $\alpha$ 
is marginally irrelevant~\cite{neto2}.  When the exchange of 
the transverse part of the vector potential
is taken into account, the velocity $v$ saturates to that of light $c$ (i.e.,  
the Lorentz invariance is recovered) and $\alpha$ remains finite in the infrared limit. 
This leads to an intriguing non-Fermi liquid state in (2+1) dimensions~\cite{Gonzalez}.  
For the (3+1)-dimensional [(3+1)D] case, the Coulomb interaction also gives the logarithmic
enhancement of the velocity $v$ and the coupling constant $\alpha$ is
marginally irrelevant~\cite{Chakravarty,Vishwanath}. 
The disorder potential is irrelevant perturbatively, while the strong enough
disorder drives the system toward the compressible diffusive metal (CDM)~\cite{Chakravarty,Shindo}.
However, the effect of 
the transverse part of the vector potential 
in (3+1) dimensions has not yet been studied to the best of our knowledge. 

In this paper, we study the quantum critical phenomenon of
topological phase transition in (3+1) dimensions. The Coulomb interaction as well as the
transverse current-current interaction are considered.   

\section{Dirac fermions in (3+1) dimensions in absence of Lorentz invariance}

\subsection{Model}
We start with the following Lagrangian~\cite{comment}: 
\begin{align}
\mathcal{L}=
	&\bar{\psi}(\gamma^0 p_0-v\vec{\gamma}\cdot\vec{p}-m)\psi
	+\frac{1}{2}(\varepsilon \vec{E}^2-\frac{1}{\mu}\vec{B}^2) \notag \\
	&-e\bar{\psi}\gamma^0\psi A_0 -e\frac{v}{c}\bar{\psi}\gamma^\alpha\psi A_\alpha,
\end{align}
where $\alpha$ is a spatial index ($\alpha=1, 2 ,3$) and 
$\gamma^\alpha p_\alpha = -\vec{\gamma}\cdot\vec{p}$.
For the moment, we consider the critical point (i.e., $m=0$).
The renormalization of the mass $m$ will be discussed later.
The speed of light in material $c$ and 
in vacuum $c_{\text{vacuum}} = 3 \times 10^8\,\mathrm{m/s}$ 
are related through the permittivity $\varepsilon$ and the permeability $\mu$ 
by $c^2=c_\text{vacuum}^2/(\varepsilon\mu)$. 
We use a $(+ - - -)$ metric.
The electric field and magnetic field are represented in terms of the photon field 
$A_\mu$ as
\[
\vec{E}=-\frac{1}{c}\frac{\partial\vec{A}}{\partial t}
-\vec{\nabla} A_0, 
\ \ 
\vec{B}=\frac{1}{c}\vec{\nabla} \times \vec{A}.
\]
The electron propagator $G_0(p)$, the photon propagator $D^{\mu\nu}_0(p)$ 
and the vertex are given by
\begin{gather}
G_0(p) = \frac{i}{\gamma^0 p_0 +v \gamma^\alpha p_\alpha +i0}
	, \\
D^{\mu\nu}_0(q) = \frac{-ig^{\mu\nu}}{\varepsilon(q_0^2-c^2q_\alpha^2)+i0}, \\
\text{vertex} = -ie\gamma^0 \mathrm{\ or\ } -ie\frac{v}{c}\gamma^\alpha.
\end{gather}
Here we employ the Feynman gauge because physical quantities are independent 
of gauge choice.

\subsection{Perturbative renormalization group analysis}
Calculations are performed by using dimensional regularization
not to violate the gauge invariance of the theory.
We set the space-time dimension $d=4 -\epsilon$ to regularize divergences.
The self-energy $\Sigma(p)$, polarization $\Pi^{\mu\nu}_2(q)$, 
and the vertex correction $\delta\Gamma^\mu(p',p)$ all to one-loop order 
(Fig.~\ref{fig:diagram}) are obtained as follows:  
\begin{figure}
\centering
\includegraphics[width=\hsize]{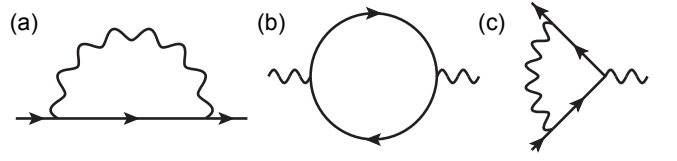}
\caption{Feynman diagrams considered here: (a) self-energy, (b) polarization, 
(c) vertex.}
\label{fig:diagram}
\end{figure}

\begin{align}
\label{eq:self-energy}
\Sigma(p) =& 
	\frac{e^2/\varepsilon}{4\pi^2\epsilon}\frac{1}{(c+v)^2c}
	\left[1-3\left(\frac{v}{c}\right)^2\right]\gamma^0p_0 \notag \\
	&+ \frac{e^2/\varepsilon}{12\pi^2\epsilon}\frac{2c+v}{(c+v)^2cv}
	\left[1+\left(\frac{v}{c}\right)^2\right]v\vec{\gamma}\cdot\vec{p},
\end{align}
\begin{equation}
\Pi^{\mu\nu}_2(q) 
    = (q^2g^{\mu\nu}-q^\mu q^\nu) 
	\left(\frac{v}{c}\right)^{2-\delta_{\mu 0}-\delta_{\nu 0}} \frac{1}{v^3}
	\Pi_2(q),
\end{equation}
where
\[
\Pi_2(q)=-\frac{e^2}{6\pi^2\epsilon}+O(\epsilon^0),
\]
and
\begin{subequations}
\label{eq:vertex}
\begin{align}
\delta\Gamma^0(0,0)&= -\frac{e^2/\varepsilon}{4\pi^2\epsilon}\frac{1}{(c+v)^2c}
	\left[ 1-3\left(\frac{v}{c}\right)^2\right] \gamma^0, \\
\delta\Gamma^\alpha(0,0)&= \frac{e^2/\varepsilon}{12\pi^2\epsilon}
	\frac{2c+v}{(c+v)^2cv}
	\left[ 1+\left(\frac{v}{c}\right)^2\right] \frac{v}{c} \gamma^\alpha.
\end{align}
\end{subequations}
Comparing the result of the vertex correction Eq.~\eqref{eq:vertex} 
with the self-energy Eq.~\eqref{eq:self-energy}, 
we can confirm that the Ward-Takahashi identity is satisfied to one-loop order.

\begin{figure}
\centering
\includegraphics[width=0.9\hsize]{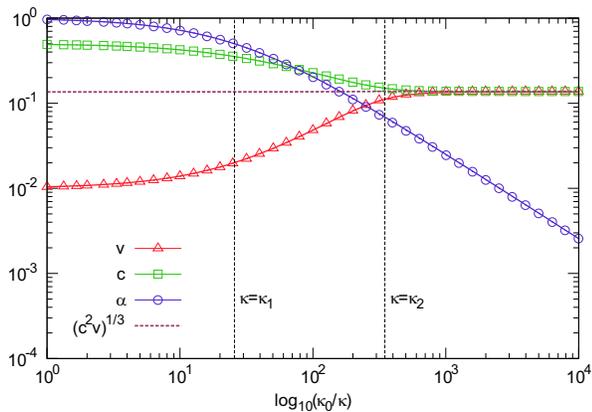}
\caption{(Color online) 
Solution to the RG equations for $v$, $c$, and $\alpha$. 
We set the initial values $v_0=0.01$, $c_0=0.5$, and $\alpha_0=1$. 
For $v$, $c$, and $\alpha$, the analytic solutions (solid lines) show 
a very good agreement with the numerical solutions (points). 
The dashed line for $c^2v$, obtained numerically, is almost constant for 
all momentum scale. }
\label{fig:rg1}
\end{figure}

The diverging quantities appearing through the calculation of the one-loop diagrams 
are absorbed by rescaling some quantities.
We can write the renormalized Lagrangian in the form
\begin{align}
\mathcal{L}=
	&\bar{\psi}(Z_{2t}\gamma^0 k_0+Z_{2s}v\gamma^\alpha p_\alpha)\psi
	+\frac{1}{2} (Z_{3e} \varepsilon\vec{E}^2 -Z_{3m} \frac{1}{\mu}\vec{B}^2)
	 \notag \\
	&-eZ_{1t}\bar{\psi}\gamma^0\psi A_0
	-eZ_{1s}\frac{v}{c}\bar{\psi} \gamma^\alpha \psi A_\alpha.
\end{align}
Then we obtain the following RG equations using a momentum scale $\kappa$:
\begin{gather}
\kappa\frac{dv}{d\kappa} = -\frac{e^2/\varepsilon}{6\pi^2}\frac{1}{(c+v)^2}
	\left[ 1+2\left( \frac{v}{c} \right)+\left( \frac{v}{c} \right)^2 
	-4\left( \frac{v}{c} \right)^3 \right] , \label{rg_v} \\
\kappa\frac{dc}{d\kappa} = \frac{e^2/\varepsilon}{12\pi^2}
	\frac{c^2-v^2}{c^3 v} , \label{eq:rg_c} \\
\kappa\frac{d(e^2/\varepsilon)}{d\kappa} =  \frac{(e^2/\varepsilon)^2}{6\pi^2}
	\frac{1}{c^2 v}. \label{eq:rg_e2} 
\end{gather}
The coupling constant $\alpha$ is defined by $\alpha=e^2/(\varepsilon c^2 v)$.
The numerical solutions for the RG equations are shown 
in Fig.~\ref{fig:rg1} for the initial (bare) values of $v_0=0.01$, $c_0=0.5$, 
and $\alpha_0 = 1$.

The result shows some important features.
First,  we can see that the quantity $c^2 v$ is almost constant 
for all momentum scales and remains $c_0^2 v_0$. 
This fact enables the approximate but accurate analysis of the 
scaling functions as described below.
Second, the speed of electron $v$ and that of photon $c$ approach to 
the same value $c_{\infty} = (c_0^2 v_0)^{1/3} $ in the infrared (IR) limit.
Third, the coupling constant $\alpha$ becomes small in the IR region, which 
justifies our perturbative RG analysis. Therefore, the quantum critical phenomenon 
of 3D topological insulator is an ideal laboratory to study the QED in a solid, even though 
the Lorentz invariance is broken to a large extent in the original (bare) Lagrangian.

\subsection{Analytic solutions}
Now we study the solution to the RG equation in more detail.  
The approximate relation $c^2 v = c_0^2 v_0$ makes the
analysis much easier, and we can obtain the analytic solution.
By replacing $c^2 v$ by $c_0^2 v_0$, the RG equation for $e^2/\varepsilon$ 
[Eq.~\eqref{eq:rg_e2}] is exactly the same as in the conventional QED. 
Therefore, the RG equation for the coupling constant $\alpha$ is  
\begin{equation}
\kappa\frac{d \alpha}{d\kappa}=\frac{\alpha^2}{6\pi^2},
\end{equation}
and its solution is obtained as
\begin{equation}
\alpha(\kappa)
=\frac{\alpha_0}{1+\dfrac{\alpha_0}{6\pi^2}\ln\left(\dfrac{\kappa_0}{\kappa}\right)},
\label{eq:alpha}
\end{equation}
where $\kappa_0$ is the momentum cutoff.
This approximate solution fits the numerical solution very well as shown 
in Fig.~\ref{fig:rg1}.
The precision of the analytic solution is discussed in Appendix.

With $\alpha(\kappa)$ being obtained, one can solve
the RG equation \eqref{eq:rg_c} for $c$ as
\begin{equation}
c^6(\kappa ) - c_0^4 v_0^2 = (c_0^6 - c_0^4 v_0^2 ) 
\biggl[ 1+\dfrac{\alpha_0}{6\pi^2}\ln\left(\dfrac{\kappa_0}{\kappa}\right) \biggr]^{ - 3}.
\label{eq:c}
\end{equation}
and $v(\kappa)= c_0^2v_0/[c(\kappa)]^2$.
These analytic solutions are again compared with the numerical solutions in 
Fig.~\ref{fig:rg1}, and a good agreement is obtained.

Here we can identify the two momentum scales, $\kappa_1$ and $\kappa_2$.
$\kappa_1$ is defined as the scale where the renormalization effect becomes
appreciable, [i.e., 
$\frac{\alpha_0}{6\pi^2}\ln(\frac{\kappa_0}{\kappa_1}) \cong 1$].
The second one $\kappa_2$ is  defined as $c(\kappa_2) \cong v(\kappa_2)$ (i.e., the
two velocities approaches to each other). These two scales are estimated as
\begin{subequations}
\begin{align}
\kappa_1=& \exp\biggl[-\frac{6\pi^2}{\alpha_0}\biggr]\kappa_0, \\
\kappa_2 =& \exp\left[-\frac{6\pi^2}{\alpha_0} 
	\biggl(\frac{c_0}{v_0}\right)^{2/3} \biggl]\kappa_0,
\end{align}
\end{subequations}
and $\kappa_2 \ll \kappa_1 \ll \kappa_0$, 
assuming $\alpha_0/(6 \pi^2) \ll 1$ and $v_0/c_0 \ll 1$.
These two momenta separate the three regions: 
(i) perturbative region $\kappa_1\ll k \ll \kappa_0$, 
the renormalization effect is small and perturbative; 
(ii) non-relativistic scaling region $\kappa_2 \ll k \ll \kappa_1$,  
the renormalization effect is large, while 
$c(\kappa) \gg v(\kappa)$ still holds; and 
(iii) relativistic scaling region $ k \ll \kappa_2$, $c \cong v$ 
and the Lorentz invariance is recovered. 

\begin{figure} 
\centering
\includegraphics[width=0.9\hsize]{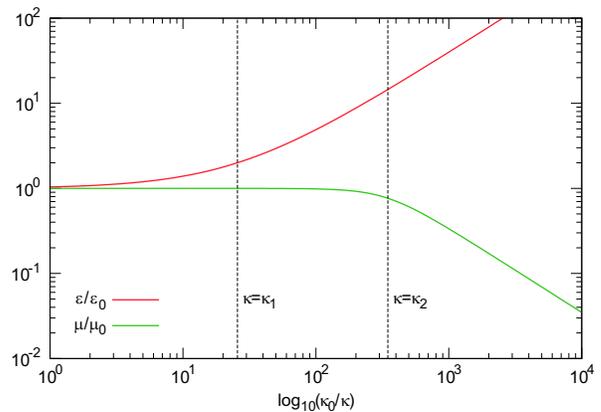}
\caption{(Color online) 
Analytic solutions to the RG equations for the permittivity $\varepsilon$ 
and the permeability $\mu$. The characteristic momentum scales are different 
for $\varepsilon$ and $\mu$. }
\label{fig:emf}
\end{figure}

\subsection{Electromagnetic properties}
Let us discuss the permittivity $\varepsilon(\kappa)$ and 
the permeability $\mu(\kappa)= 1 + 4 \pi \chi$ ($\chi$: magnetic susceptibility).
The analytic solutions obtained from Eq.~\eqref{eq:rg_e2} and 
$\mu=1/(\varepsilon c^2)$ are shown in Fig.~\ref{fig:emf}.
The momentum scale $\kappa$ can be regarded as the temperature $T$
by $T \cong v(\kappa) \kappa$. 
As noted above, the velocity $v(\kappa)$ is the function of the momentum scale,
hence the energy dispersion $E(k) = v(k) k$ is a nonlinear function of $k$.
From the definition of $\kappa_1$ and $\kappa_2$, $v(\kappa_1) \cong v_0$ and
$v(\kappa_2) \cong c_{\infty}$, and the corresponding temperatures are 
estimated as  $T_1 = T(\kappa_1) \cong v_0 \kappa_1$ and 
$T_2 = T(\kappa_2) \cong c_{\infty} \kappa_2$. 
In Fig.~\ref{fig:emf}, it can be seen that the permittivity $\varepsilon(\kappa)$ 
grows logarithmically below $T_1$ while the permeability $\mu(\kappa)$ decreases
below $T_2$. 
The orbital magnetic susceptibility $\chi$ without the electron-electron interaction 
logarithmically diverges as a function of $T$, 
but in our analysis, the logarithmic divergence 
is canceled due to the renormalization of $v$.
These contrasting behaviors of $\varepsilon$ and $\mu$ facilitate
the identification of $T_1$ and $T_2$ experimentally.
In the zero temperature limit $\varepsilon$ diverges while
$\mu$ goes to zero, i.e., the perfect diamagnetism $\chi = - 1/(4 \pi)$ is accomplished.

\subsection{Spectral function}    
For the physical interpretation of the electron Green's function, we should consider 
the self-energy in Coulomb gauge as discussed above. 
In Coulomb gauge~\cite{Adkins}, the electron self-energy is 
\begin{align}
\Sigma(p) =& -\frac{e^2/\varepsilon}{2\pi^2\epsilon}\frac{v^2}{c^3(c+v)^2}
	\gamma^0p_0 
	+\frac{e^2/\varepsilon}{6\pi^2\epsilon}\frac{1}{v(c+v)^2} \notag \\
	&\times \left[ 1+2\left( \frac{v}{c} \right) +\left( \frac{v}{c} \right)^2
	- \left( \frac{v}{c} \right)^3 \right] v\vec{\gamma}\cdot\vec{p}.
\end{align}
In principle, ARPES can measure the energy dispersion $E(k) = v(\kappa= k) k$, 
which shows crossovers at $\kappa_1$ and $\kappa_2$. 
For the spectral function of electrons, 
the electron field renormalization is required, which is given by
\begin{equation}
\gamma_2(v,c,e^2/\varepsilon;\kappa) = \frac{1}{2}\kappa\frac{d \ln Z_{2t}}{d \kappa} = 
	\frac{e^2/\varepsilon}{4\pi^2}\frac{v^2}{c^3(c+v)^2}.
\end{equation}
From the Callan-Symanzik equation, the electron Green's function is modified by 
the momentum-scale dependent functions $v(k)$, $\alpha(k)$, and $\gamma_2(k)$, 
then we obtain
\begin{equation}
G(\vec{k},\omega)=\frac{\mathcal{G}(\alpha(k))}{\omega^2-v^2(k)\vec{k}^2}
	\exp\left[ 2\int_\Lambda^k d\ln\left( \frac{k'}{\Lambda} \right) 
	\gamma_2(\alpha) \right],
\end{equation}
where $k'=(\omega',v\vec{k'})$ is a four-momentum, $\Lambda$ is the energy cutoff, 
and $\mathcal{G}$ is a function determined from a perturbative renormalization 
calculation. 

\begin{figure}
\centering
\includegraphics[width=0.9\hsize]{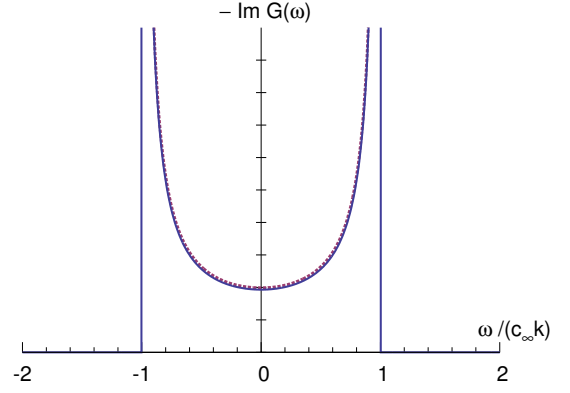}
\caption{(Color online) 
Frequency dependence of the density of states in region (iii) with $\alpha_0=1$.
The vertical axis is in linear but arbitrary scale. The solid line denotes the result with 
the perturbative correction in $\mathcal{G}$, while the dashed line depicts 
the result for $\mathcal{G}=1$.}
\label{fig:dos}
\end{figure}

In region (i), $\gamma_2=0$, so the Green's function is unchanged.
In region (ii), $\kappa$ dependence of $\gamma_2$ is rather complicated to calculate 
$G(\vec{k},\omega)$, so we only consider the relativistic scaling region (iii), 
where $v$ approaches $c$ and the original QED regime is applicable. 
When we put $c=v=c_\infty$, $\gamma_2(k)$ is expressed as
\begin{equation}
\gamma_2(k)=\frac{\alpha(k)}{16\pi^2}=
	\frac{\alpha_0}{16\pi^2}\frac{1}{1+\dfrac{\alpha_0}{6\pi^2}
	\ln\left(\dfrac{\Lambda}{k}\right)},
\end{equation}
and the perturbative correction for $\mathcal{G}$ is 
\begin{equation}
\mathcal{G}(\alpha(k)) = 1+ \frac{\alpha(k)}{16\pi^2}
	\ln\left(\frac{e^\gamma}{4\pi}\right) + \mathcal{O}(\alpha^2).
\end{equation}
Then, the Green's function becomes
\begin{equation}
G(\vec{k},\omega)=\frac{\mathcal{G}(\alpha(k))}{\omega^2-c_\infty^2\vec{k}^2}
	\frac{1}{\biggl[ 1+\dfrac{\alpha_0}{12\pi^2}
	\ln\biggl(\dfrac{\Lambda^2}{\omega^2-c_\infty^2\vec{k}^2}\biggr)\biggr]^{3/4}}.
\end{equation}
By substituting $\omega$ with $\omega +i0$, 
the imaginary part of the Green's function
$-\mathrm{Im}G(\vec{k},\omega)$ gives the electron spectral function.
The electron spectral function has finite value for $|\omega|<|\vec{k}|$, while
$-\mathrm{Im}G(\vec{k},\omega)=0$ outside that region. 
As depicted in Fig.~\ref{fig:dos}, the perturbative correction for $\mathcal{G}$ 
gives very small contribution, so we put $\mathcal{G}=1$ in the following analysis. 
When the bare coupling constant $\alpha_0^2$ is small enough, the spectral function 
has the approximate form
\begin{align}
&-\mathrm{Im}G(\vec{k},\omega) \notag \\
\sim & a\delta(\omega^2-c_\infty^2\vec{k}^2) + 
	\frac{\alpha_0}{32\pi c_\infty |\vec{k}|}
	\biggl( \frac{1}{c_\infty|\vec{k}|-\omega}+\frac{1}{c_\infty|\vec{k}|+\omega} \biggr),
\end{align}
where the residue $a$ is a constant determined from the sum rule.
The $\delta$ function peak with finite $a$ means that the system remains a
Fermi liquid in sharp contrast to the (2+1)D case~\cite{Gonzalez},
while the continuum state for $|\omega| < c_{\infty} k$ comes from the interaction 
as shown in Fig.~\ref{fig:dos}.

\subsection{Energy gap}
Up to now, we have focused on the critical point ($m=0$), but the mass $m$ 
is a relevant parameter. Experimentally, the bare mass $m_0$ can be controlled by
the concentration $x$ or by pressure $P$~\cite{Hasan, Sato}.  
The RG equation for mass $m(\kappa)$ is
\begin{equation}
\kappa\frac{d m(\kappa)}{d\kappa}=-\frac{3\alpha(\kappa)}{8\pi^2}m.
\end{equation}
Then, the mass at momentum scale $p$ is 
\begin{equation}
\label{eq:mass}
m(p)=m(\Lambda) 
	\left[ 1+\frac{\alpha_0}{6\pi^2}\ln \left( \frac{\Lambda}{p} \right) \right]^{9/4}.
\end{equation}
When we neglect the weak singularity with $\log\log m_0$, 
the solution to Eq.~\eqref{eq:mass} is given by 
$m=m_0 [ 1+\frac{\alpha_0}{6\pi^2}\ln ( \frac{\Lambda}{m_0} ) ]^{9/4}$, 
which describes the critical behavior of the gap as a function of
$m_0 \propto (x - x_c)$ or $m_0 \propto (P-P_c)$ with $x_c$ $(P_c)$ being
the critical concentration (pressure).  

\section{Discussions and Summary}
Now we discuss the relevance of the present results to the real systems. 
The velocity $v_0$ is estimated at $v_0 \cong 10^6\,\mathrm{m/s}$ 
from the ARPES measurement of the energy dispersion~\cite{Hasan}, 
hence $c_{\text{vacuum}}/v_0 \cong 300$.
As the dielectric constant, we take the typical value
$\varepsilon_0 \cong 10^2$ for BiSb alloys~\cite{monopole}.
Since $c_0 = c_{\text{vacuum}}/\sqrt{\varepsilon_0}$, 
$c_0/v_0 \cong 30$, $(c_0/v_0)^{2/3} \cong 10$, and 
$\alpha_0 = (c_{\text{vacuum}}/v_0) \times 1/137 \cong 3$ are obtained.
These values give the estimates for $\kappa_1 \cong 10^{-8} \kappa_0$
and $\kappa_2$ being extremely small. 
Unfortunately, it would be difficult
to observe the effect of electron-electron interaction and the scaling behavior 
at the experimentally accessible temperature in the materials at hand.
However, there are many candidates for the correlated topological insulators 
recently proposed and partly synthesized
~\cite{Shitade,Giniyat,NIO2,NIO4,NIO6,JJ,NIO1,NIO3,NIO5,NIO7,pyrochlore,Okamoto}.
The smaller value of $v_0$ rapidly (exponentially) increases $\kappa_1$, 
which gives the clue to look for the appropriate materials to study the scaling
behavior of the quantum criticality. 

In summary, we have studied the (3+1)D Dirac electrons coupled to 
electromagnetic field as the model for the quantum critical phenomenon 
of the transition between topological insulator and trivial insulator.
The RG equations are derived and the two scaling regions are identified, 
(i.e., the nonrelativistic and relativistic scaling regions). 
The Lorentz invariance is recovered in the latter case. 
The physical properties such as the the permittivity, the permeability, 
and the electron spectral function have been discussed based on the RG equations.


\begin{acknowledgements}
The authors acknowledge the fruitful discussion with T. Hatsuda. 
This work is supported by a Grant-in-Aid for Scientific Research
(Grant No. 24224009)
from the Ministry of Education, Culture,
Sports, Science and Technology of Japan, Strategic
International Cooperative Program (Joint Research Type)
from Japan Science and Technology Agency, and Funding
Program for World-Leading Innovative RD on Science and
Technology (FIRST Program).
\end{acknowledgements}

\appendix*
\section{Precision of the analytic solutions}

In the appendix, we consider the precision of the analytic solutions to 
the RG equations, especially the validity of the relation $c^2v=\mathrm{const.}$
In this section, we define $x=1-v/c$, $y=c^2 v$. The RG equations for 
$x$, $y$ and $e^2/\varepsilon$ are
\begin{gather}
\kappa\frac{dx}{d\kappa} = \frac{e^2/\varepsilon}{12\pi^2}\frac{1}{y}f(x), 
	\label{eq:rg_x} \\
\kappa\frac{dy}{d\kappa} = -\frac{e^2/\varepsilon}{6\pi^2}g(x), \label{eq:rg_y} \\
\kappa\frac{d(e^2/\varepsilon)}{d\kappa} = \frac{(e^2/\varepsilon)^2}{6\pi^2}\frac{1}{y},
	\label{eq:rg_e}
\end{gather}
where
\begin{gather}
f(x) = \frac{x(1-x)(24-34x+12x^2-x^3)}{(2-x)^2}, \\
g(x) = \frac{x^2(1-x)^2}{(2-x)^2}.
\end{gather}
From Eqs. \eqref{eq:rg_y}, \eqref{eq:rg_e}, we obtain
\begin{equation}
\label{eq:diff}
\frac{e^2/\varepsilon}{y}\frac{dy}{d(e^2/\varepsilon)} = -g(x).
\end{equation}
We assume $v \leq c$, i.e. $0 \leq x \leq 1$, so that 
$0 \leq g(x) \leq 17-12\sqrt{2} \cong 0.03$.
The RHS of Eq. \eqref{eq:diff} is small, so the relative difference 
of $y$ is
\begin{equation}
\frac{y(\kappa)}{y_0} 
\sim \left[ \frac{(e^2/\varepsilon)(\kappa)}{e_0^2/\varepsilon_0} \right]^{-g(x)}.
\end{equation}
The maximum value $g(x) \cong 0.03$ is rarely observed.
Thus, $y=c^2 v$ can be regarded as a constant to a large extent.



\begin{thebibliography}{99}
\bibitem{comment}
We have dropped the $\theta$ term [i.e., $\theta\vec{E}\cdot\vec{B}$ 
($\theta = \pm \pi$) in the action], which should be present in the topological 
insulator phase.  This term, however, can be transformed into the surface term, 
and the sign of $\theta$ is determined by the time-reversal symmetry breaking 
on the surface.  The topological magneto-electric effect is derived from this term, 
but this is beyond the scope of this present paper, where only the permittivity 
$\varepsilon$ and the magnetic permeability $\mu$ are discussed.
\end{thebibliography}
\end{document}